\documentclass[aps,twocolumn,prr]{revtex4-1}
\usepackage{amsmath}
\usepackage{braket}
\usepackage{cancel}
\usepackage{mathtools} 
\usepackage{mathbbol}
\usepackage{siunitx} 
\usepackage[pdftex]{graphicx}
\newcommand{\od}{\omega_\mathrm{d}}
\newcommand{\ed}{E_\mathrm{d}}
\begin{document}
\title{Detecting light-induced Floquet band gaps of graphene via trARPES}
\author{Lukas Broers$^{1,2}$}
\author{Ludwig Mathey$^{1,2,3}$}
\affiliation{
$^{1}$Center for Optical Quantum Technologies, University of Hamburg, 22761 Hamburg, Germany\\
$^{2}$Institute for Laser Physics, University of Hamburg, 22761 Hamburg, Germany\\
$^{3}$The Hamburg Center for Ultrafast Imaging, Luruper Chaussee 149, 22761 Hamburg, Germany}
\begin{abstract}
We propose a realistic regime to detect the light-induced topological band gap in graphene via time-resolved angle-resolved photoelectron spectroscopy (trARPES), that can be achieved with current technology.
The direct observation of Floquet-Bloch bands in graphene is limited by low-mobility, Fourier-broadening, laser-assisted photoemission (LAPE), probe-pulse energy-resolution bounds, space-charge effects and more. 
We characterize a regime of low driving frequency and high amplitude of the circularly polarized light that induces an effective band gap at the Dirac point that exceeds the Floquet zone.  
This circumvents limitations due to energy resolutions and band broadening. 
The electron distribution across the Floquet replica in this limit allow for distinguishing LAPE replica from Floquet replica. 
We derive our results from a dissipative master equation approach that gives access to two-point correlation functions and the electron distribution relevant for trARPES measurements.
\end{abstract}
\maketitle

Floquet engineering constitutes a novel approach to control material properties via light \cite{Basov17,Lindner11,Kitagawa2011,Bukov15}. 
A prominent example is the proposed light-induced topologically insulating state of monolayer graphene \cite{oka,Usaj14}. 
The resulting anomalous Hall effect in this system has been observed experimentally \cite{mciver} and explained as a geometric-dissipative effect \cite{nuske} in accordance with Floquet theory.
Meanwhile, time-resolved angle-resolved photoemission spectroscopy (trARPES) is established as the prime method for resolving dynamical changes in effective band structures of solid state systems \cite{Kutnyakhov20,Cacho12,Peli20,Rohde16,Eich14,Puppin19,Nicholson}.
Experimental trARPES setups are constantly improving and being used for investigating the dynamical electronic processes in two-dimensional Dirac materials such as graphene \cite{Gierz14,Ulstrup15,Gierz17,Aeschlimann20,Gierz21}, $\mathrm{W}\mathrm{Se}_2$ \cite{Dong21,Maklar20,Liu17,Aeschlimann20,Gierz21} or $\mathrm{Bi}_2\mathrm{Se}_3$ \cite{Gierz15b,Gedik2,Soifer19,Wang11}.
Approaches related to observing pseudospin textures in ARPES have been discussed in Refs.~\cite{Sentef15,Schuler20,Beaulieu20}.
In $\mathrm{Bi}_2\mathrm{Se}_3$, the Floquet replica of electronic bands have been observed using trARPES setups \cite{Gedik}.
However, the direct observation of both the replica and the topological gap at the Dirac point are met with intricate challenges in graphene and remain unachieved to date.

\begin{figure}
\centering
\includegraphics[trim={0 1cm 0 0},clip,width=1.0\linewidth]{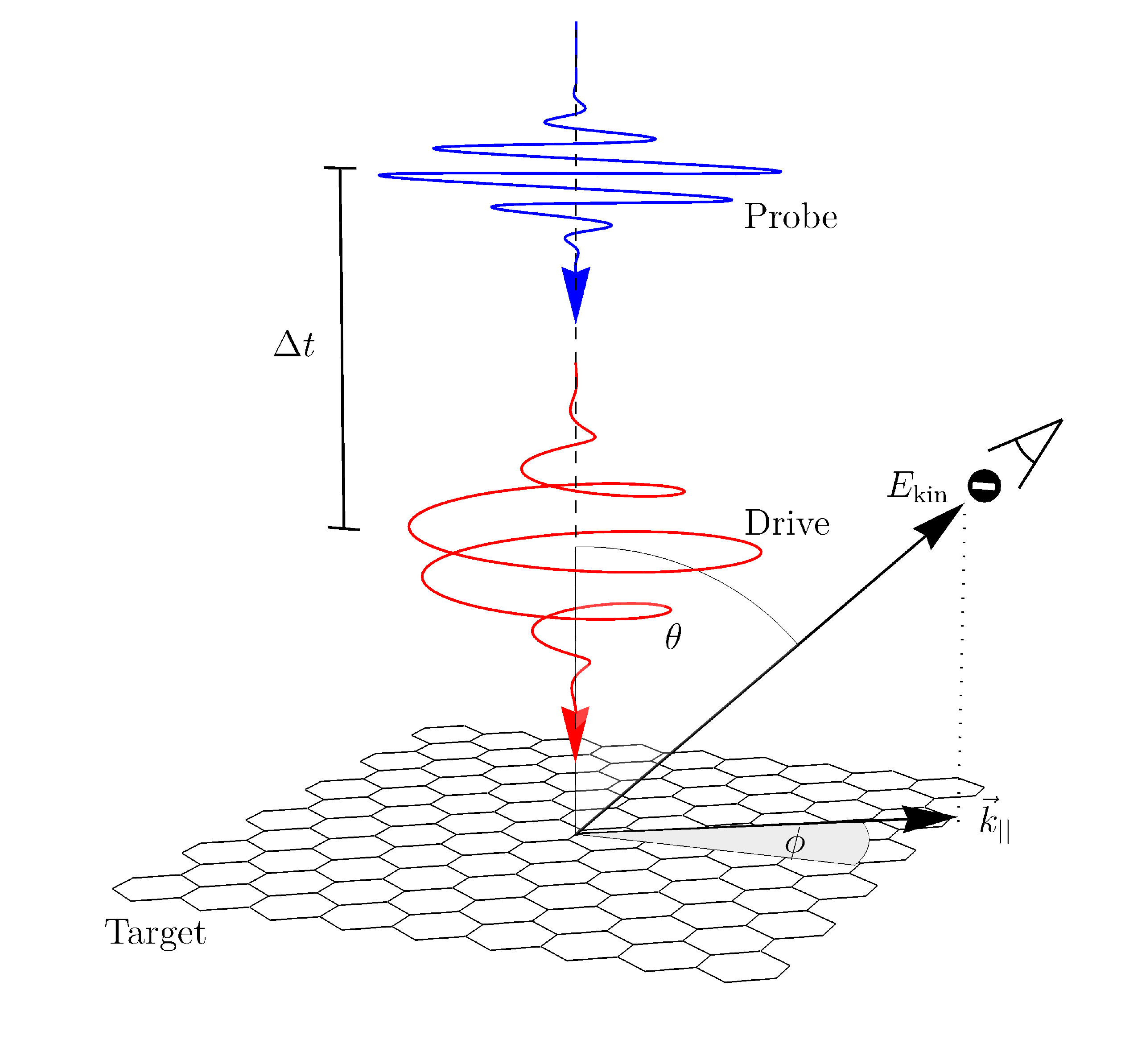}
\caption{
An illustration of the trARPES principle. 
A circularly polarized infrared drive pulse (red) hits the graphene target from a perpendicular direction, exciting a transient state in the illuminated region.
After a time delay of $\Delta t$, an ultraviolet probe pulse (blue) hits the target and excites photoelectrons. 
A given electron leaves the target with a kinetic energy $E_\mathrm{kin}$ at an inclination $\theta$ to the graphene target and at an azimuthal angle $\phi$.
The momentum component $\vec{k}_{||}$ parallel to the target is that of the pre-probe transient state of the electron bound in the graphene layer. 
}
\label{trarpes}
\end{figure}

In this work we determine the regime of trARPES measurements for observing the topological band gap at the Dirac point of irradiated graphene.
We propose to perform these measurements in the regime of low driving frequencies and high driving field strengths.
In this previously unexplored regime the dominant Floquet-Bloch band occupations are spaced farther apart than the driving frequency.
They are therefore outside of the first Floquet zone.
We propose to detect this light-induced energy gap beyond the Floquet zone in experiment, because the different spectral features are well resolved in this regime.
We discuss the dependence of our predictions on the system parameters, specifically how they affect the systematic limitations of the energy resolution of photoemission spectroscopy.
These parameters include the driving frequency and field strength, which determine the Floquet-Bloch band structure, 
the dissipation coefficients that broaden the band signals, and the pulse lengths of drive and probe lasers. 
For the pulse lengths we point out a desirable regime with sufficient energy resolution and high enough repetition rates.
These repetition rates are necessary to avoid undesired space-charge effects, where the photoemitted electrons interact and affect each others trajectories \cite{Hellmann12, Hellmann, Oloff16,Maklar20}.
In particular, it is possible to distinguish laser-assisted photoemission (LAPE) replica from Floquet replica within the gap at the Dirac point in our suggested regime.
 
\begin{figure*}
\centering
\includegraphics[width=1.0\linewidth]{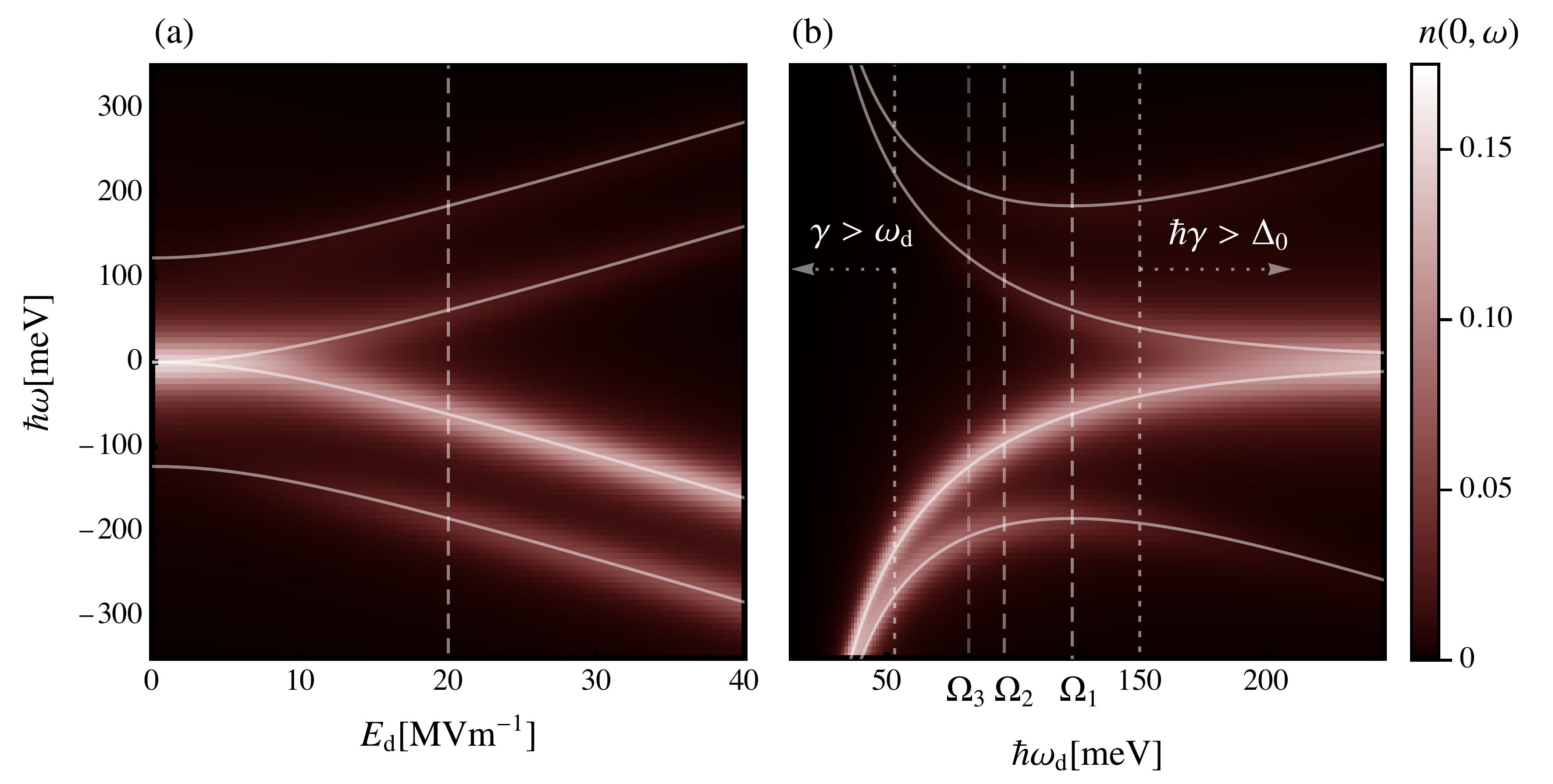}
\caption{
The electron distribution at the Dirac point $n({\bf k}=0,\omega)$ for zero-delay ($\Delta t = 0$) 
as a function of the driving field strength at the driving frequency $\od=2\pi\times 29.8\si{\tera\hertz}\approx123\si{\milli\electronvolt}\hbar^{-1}$ (a) 
and as a function of the driving frequency at the driving field strength $\ed = 20\si{\mega\volt\per\meter}$ (b).
The solid lines show the expected Floquet energies $\pm\sqrt{(ev_F\ed/\od)^2 +(\hbar\od/2)^2}\pm\hbar\omega_\mathrm{d}/2$. 
The dashed lines indicate the field strength at which $\Delta_0=\hbar\od$ (a) and the driving frequencies $\Omega_1$, $\Omega_2$ and $\Omega_3$ (b).
The frequencies $\Omega_{m=1,2,3}$ are defined in Eq.~\ref{wdcond}. The examples shown in Figs.~\ref{fullnkw} and \ref{delayscans} also use these frequencies.}
\label{wdscan}
\end{figure*}

We consider a single layer of graphene irradiated by a circularly polarized infrared laser from a perpendicular direction.
We consider a laser pulse with a temporal Gaussian envelope of pulse length $\tau_\mathrm{d}$. 
The pulse length is assumed to be much longer than the driving period, so that it induces Floquet-Bloch states, that vary with the envelope function of the pulse. 
The graphene sample is probed by a tunable extreme ultraviolet (XUV) laser pulse from the same direction.
It has a shorter pulse length and excites photoelectrons out of the driven graphene over a time-span during which the driving intensity is approximately constant.
This is necessary for resolving the time-dependent Floquet-Bloch bands which are sensitive to the driving amplitude. 
This is considered the standard approach to trARPES experiments \cite{ARPESreview} and is illustrated in Fig.~\ref{trarpes}.
The emitted photoelectrons corresponding to a probe frequency $\omega_\mathrm{p}$ have the kinetic energy 
\begin{equation}
    E_\mathrm{kin} = \hbar \omega_\mathrm{p} - E_\mathrm{b} - \Phi,
\end{equation}
where $E_\mathrm{b}$ is the binding energy and $\Phi$ is the work function of the material, which is the energy required to remove the electron from the graphene.
In addition, a photoelectron has the momentum $\vec{k}$ with components
\begin{equation}
\vec{k}_{||} = \sqrt{2mE_\mathrm{kin}}\hbar^{-1}\sin(\theta) (\cos(\phi),\sin(\phi))^\mathrm{T}
\end{equation}
parallel to the graphene layer.
$m$ is the electron mass, and $\theta$ and $\phi$ are the inclination and azimuthal angles of the momentum with respect to the graphene layer.
Measuring the photoelectron counts at these angles and energies gives access to the time-resolved Floquet-Bloch bands.
The momentum $\vec{k}_{||}$ is the electron momentum prior to the excitation process.
For simplicity this is denoted as ${\bf k}$ in the following.

We describe the electron dynamics in graphene with the Hamiltonian 
\begin{equation}
H(t) = \sum_{\bf k} c_{\bf k}^\dagger h({\bf k},t) c_{\bf k},
\end{equation}
where $c_{\bf k}=(c_{{\bf k},A},c_{{\bf k},B})^\mathrm{T}$. 
The $c^{(\dagger)}_{{\bf k},i}$, with $i\in\{A,B\}$, are the annihilation (creation) operators of the graphene sublattices.
The single-particle Hamiltonian is given by
\begin{equation}
h({\bf k},t) = \hbar v_F (q_x(t) \sigma_x+q_y(t) \sigma_y)
\end{equation}
with 
\begin{align}
q_x(t) &= k_x+\frac{e}{\hbar}A_\mathrm{d}(t)\sin(\omega_\mathrm{d}t)\\
q_y(t) &= k_y+\frac{e}{\hbar}A_\mathrm{d}(t)\cos(\omega_\mathrm{d}t),
\end{align}
where $v_F\approx 10^6 \si{\meter\per\second}$ is the Fermi velocity and $k_x=|{\bf k}|\cos(\phi)$ and $k_y=|{\bf k}|\sin(\phi)$ are the momentum components. $\sigma_i$ are the Pauli matrices.
The pulsed vector potential amplitude is given by
\begin{equation}
A_\mathrm{d}(t) = \frac{E_\mathrm{d}}{\omega_\mathrm{d}} \exp\{-t^2\tau_\mathrm{d}^{-2}4\ln(2)\} ,
\end{equation}
where $\tau_\mathrm{d}$ is the driving pulse full width at half maximum (FWHM). 
For the pulse length we use the value $\tau_\mathrm{d}= 500\si{\femto\second}$ as a realistic value for driving frequencies in the range of tens to hundreds of terahertz. 

We consider a product state $\rho=\Pi_{\bf k}\rho_{\bf k}$ and evolve the system using a Lindblad-von Neumann master equation that includes dissipation.
The dissipation channels amount to dephasing, decay and electron exchange with a backgate. 
The corresponding coefficients are chosen as $\gamma_z=22.5\si{\tera\hertz}\approx (44.4\si{\femto\second})^{-1}$, $\gamma_\pm=10\si{\tera\hertz}=(100\si{\femto\second})^{-1}$ and $\gamma_\mathrm{bg}=25\si{\tera\hertz}=(40\si{\femto\second})^{-1}$.
These are the values that were demonstrated to describe the experimental realization of Ref.~\cite{mciver}, in Ref.~\cite{nuske}. 
These values also agree with the relaxation times of $20\si{\femto\second}$ to $40\si{\femto\second}$ found in Refs.~\cite{Hommelhoff1,Gierz15,Breusing9,Gierz13}, and the electron-phonon channel relaxation estimated to be on the order of $100\si{\femto\second}$ \cite{Breusing9,Gierz13,Elsaesser1}.
We include a non-zero temperature in the system by giving complementary dissipation coefficients the corresponding Boltzmann factors so that the equilibrium states realizes the desired temperature $T$.
For details of this approach, see Ref.~\cite{nuske}. 
Throughout this work we use room temperature $T=300\mathrm{K}$.

Our predictions for the trARPES measurements are based on the momentum- and energy-resolved electron distribution calculated as \cite{Freericks9}
\begin{equation}
n(k,\omega) = \int_{-t_0}^{t_0}\int_{-t_0}^{t_0} s(t_1)s(t_2)\mathcal{G}(k,t_2,t_1)\frac{e^{i\omega(t_2-t_1)}}{4t_0^2}\mathrm{d}t_2\mathrm{d}t_1
\label{arpes}
\end{equation}
with the correlator \footnote{
This expression for $\mathcal{G}(k,t_2,t_1)$ omits coherence terms such as $c^\dagger_{k,A}(t_2) c_{k,B}(t_1)$, which are responsible for the dark corridor in graphene.
}
\begin{equation}
\mathcal{G}({\bf k},t_2,t_1)=\sum_{i\in\{A,B\}}\braket{c^\dagger_{{\bf k},i}(t_2)c_{{\bf k},i}(t_1)}
\end{equation}
and the probe pulse envelope
\begin{equation}
s(t) = \exp\{-(t-\Delta t)^2 \tau_\mathrm{p}^{-2}4\ln(2)\},
\end{equation}
where $\tau_\mathrm{p}$ is the probe pulse length (FWHM) and $\Delta t$ is the delay time between the incidence of the drive and probe pulses.
For the probe length we use the value $\tau_\mathrm{p} = 100\si{\femto\second}$.
$t_0$ is the temporal integration range, for which we choose $t_0 = 3\tau_\mathrm{d}$, to support the probe pulse sufficiently well.

We choose the pulse lengths of the drive and the pump pulse to fulfill two requirements. Firstly, the probe pulse length is chosen to be short compared to the drive pulse length, so that the drive-induced dynamics is resolved. Secondly, the probe pulse length is chosen to be large compared to the driving period. 
These conditions are expressed as
\begin{align}
    \tau_\mathrm{d} \gg \tau_\mathrm{p} \gg \frac{2\pi}{\omega_\mathrm{d}}.
\end{align}
When the probe pulse length $\tau_\mathrm{p}$ and the driving period $2\pi/\od$ are comparable, Eq.~\ref{arpes} no longer resolves Floquet-Bloch bands but rather sub-driving period electron occupations.
We note that increasing the pulse lengths requires increasing the drive pulse energies, which are experimentally limited.
This also leads to a reduced repetition rate, which results in undesirable space-charge effects that greatly decrease the resolution due to electron scattering \cite{Hellmann, Hellmann12, Graf10}.
We do not include this effect in our numerics, but acknowledge that it necessitates a compromise in the pulse lengths,
which is reached with the given values of $\tau_\mathrm{p}$ and $\tau_\mathrm{d}$.

In Fig.~\ref{wdscan} (a), we show the electron distribution $n({\bf k}=0,\omega)$ at the Dirac point at zero-delay, i.e. $\Delta t = 0$.
We choose the driving frequency $\od=2\pi\times29.8\si{\tera\hertz}$, and display the electron distribution as a function of the driving field strength $\ed$.
We refer to the energy difference of the two distribution maxima that emerge at ${\bf k} = 0$ as the energy gap $\Delta_0$.
We see that this gap $\Delta_0$ grows monotonously as a function of $E_\mathrm{d}$ rather than being confined within the first Floquet zone of width $\od$, as also discussed in Ref.~\cite{broers}. 
We derive the Floquet energy gap at the Dirac point from $h({\bf k}=0,t)$ with a fixed vector potential amplitude $A_\mathrm{d}=E_\mathrm{d}/\omega_\mathrm{d}$.
Using the Rabi solution we obtain the frequency 
\begin{equation}
    \Delta_0/2 = \sqrt{(ev_F\ed/\od)^2 +(\hbar\od/2)^2}-\hbar\od/2.
    \label{delta0}
\end{equation}
In the following, we point out the most promising regime in which this gap can be detected. 
As we display in Fig.~\ref{wdscan} (a), the gap $\Delta_0$ grows with increasing driving strength $\ed$, in particular it grows beyond the Floquet zone boundary at $\od/2$. 
We propose to detect the energy gap $\Delta_0$ in this strongly driven regime in which $\hbar\od\approx\Delta_0$.
While the Floquet quasi-energies are confined to the Floquet zone, the maxima of the electron distribution continue to be shifted to higher frequencies with increasing $\ed$ so that they can be resolved despite broadening effects and energy resolution limitations.
For very large field strengths with $\Delta_0\gg \hbar\od$, the electrons will predominantly populate the lower bands at the Dirac point.
Therefore, intermediate values of $\ed$ are desirable, such that $\Delta_0\approx\hbar\od$, as we discuss throughout this paper.

In Fig.~\ref{wdscan} (b), we show the electron distribution $n({\bf k}=0,\omega)$ at zero-delay, i.e. $\Delta t = 0$, at the Dirac point for the driving field strength $\ed=20\si{\mega\volt\per\meter}$ as a function of the driving frequency $\od$.
We display the energies $\pm \sqrt{(ev_F\ed/\od)^2 +(\hbar\od/2)^2} \pm \hbar\od/2$ which reproduce the maxima of the electron distribution.  
We see the expected scaling behavior of the gap at the Dirac point $\Delta_0$ of Eq.~\ref{delta0} as well as the spacing between the nearest Floquet replica $\Delta_0+\hbar\od$.
We propose to measure the electron distribution in the regime that is given by $\od > \gamma$ and $\Delta_0 > \hbar\gamma$. $\gamma$ is given by $\gamma=\gamma_\pm/2+2\gamma_z+\gamma_\mathrm{bg}= 75\si{\tera\hertz}\approx 50\si{\milli\electronvolt}/\hbar\approx(13.3 \si{\femto\second})^{-1}$, as an overall metric for the decay rate.
This value for $\gamma$ is comparable to the coherence times of $22\si{\femto\second}$ found in Ref.~\cite{Hommelhoff2}.
With increasing $\od$ and for fixed $\ed$, $\Delta_0$ decreases. This dependence is predicted by Eq.~\ref{delta0}. 
If $\Delta_0$ is smaller than $\hbar\gamma$, the two maxima of the electron distribution are not resolved, and are not detectable via trARPES.
With decreasing $\od$, $\Delta_0$ increases and becomes easier to resolve. 
However, the driving period $2\pi/\od$ needs to be shorter than the characteristic timescale of the dissipative processes, i.e. $\od>\gamma$. 
Otherwise the picture of a close to adiabatically stirred Dirac cone in equilibrium is more appropriate than that of emerging Floquet-Bloch bands.
Long scattering times have also been connected to the visibility of $\Delta_0$ in Ref.~\cite{Aeschlimann21}.
The range of feasible driving frequencies given by these two conditions decreases for increasing $\gamma$ but increases for increasing $E_\mathrm{d}$.

\begin{figure}
\centering
\includegraphics[trim=0 0 0 0,width=1.0\linewidth]{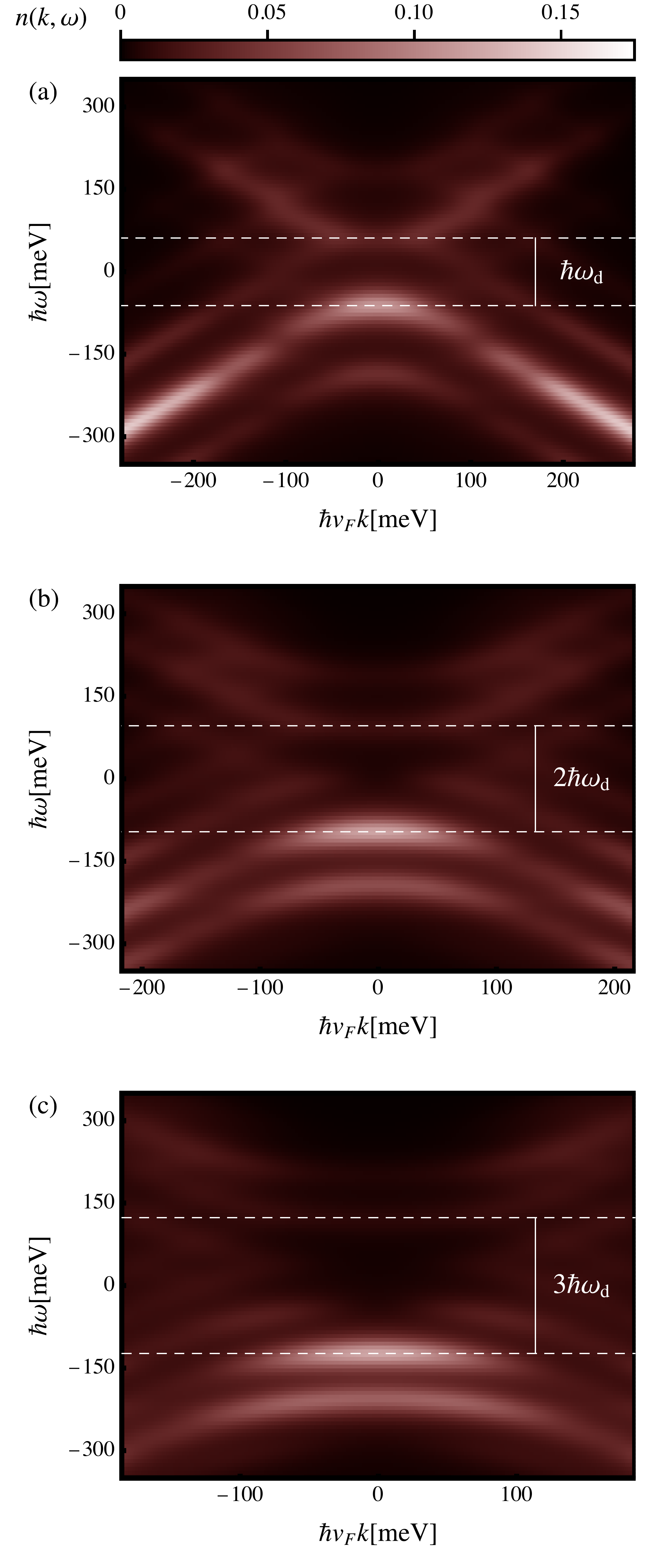}
\caption{
The electron distribution $n({\bf k},\omega)$ for zero-delay ($\Delta t = 0$) and the driving field strength $\ed=20\si{\mega\volt\per\meter}$.
The driving frequenceis are $\od=\Omega_1$ (a), $\od=\Omega_2$ (b) and $\od=\Omega_3$ (c).
The dashed lines indicate the effective gap $\Delta_0$ of size $\hbar\od$ (a), $2\hbar\od$ (b) and $3\hbar\od$ (c).
}
\label{fullnkw}
\end{figure}

\begin{figure}
\centering
\includegraphics[trim=0 0 0 0,width=1.0\linewidth]{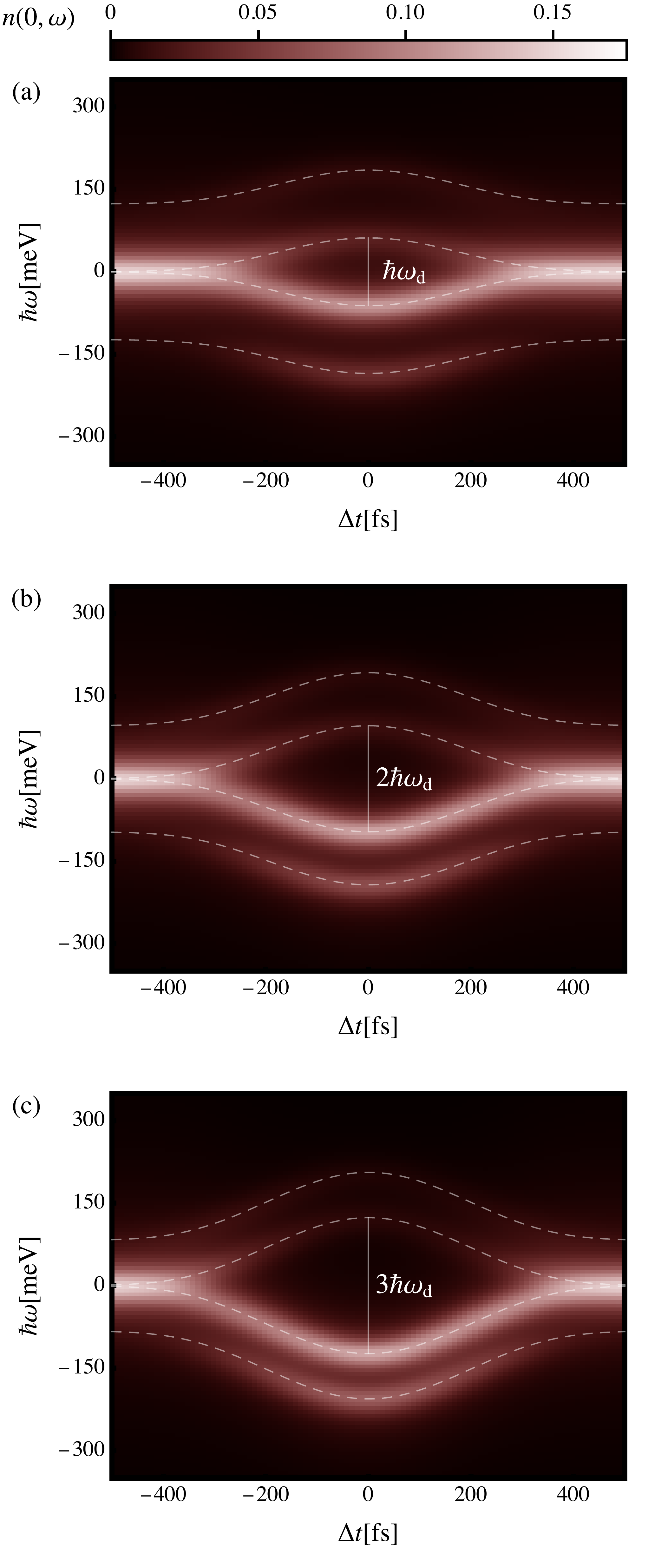}
\caption{
    The electron distribution $n({\bf k}=0,\omega)$ at the Dirac point as a function of the delay time $\Delta t$ for a peak driving field strength of $\ed=20\si{\mega\volt\per\meter}$.
    The driving frequencies are $\od=\Omega_1$ (a), $\od=\Omega_2$ (b) and $\od=\Omega_3$ (c).
    The dashed lines indicate the static Floquet energies corresponding to the driving field strengths at the center of the probe pulse.
}
\label{delayscans}
\end{figure}

In general, the $m$-photon gaps $\Delta_m$ open up at the momenta of $v_F |{\bf k}| = m \omega_\mathrm{d}/2$, with $m>0$, for small field strengths $\ed$. 
These gaps move inwards towards the Dirac point for increasing field strengths. 
Thus increasing $\ed$ increases the gap at the Dirac point but at the same time decreases the distance in momentum space to higher order gaps. 
For the driving field strength \cite{broers}
\begin{equation}
    E_\mathrm{d} = \frac{\hbar\od^2}{e v_F} \sqrt{\frac{m^2}{4}+\frac{m}{2}}
    \label{edcond}
\end{equation}
the $m$th gap is located at ${\bf k}=0$ and merges with the gap $\Delta_0$. 
The next gap $\Delta_{m+1}$ is then the gap closest to the Dirac point with its distance maximized with respect to $\ed$.
This further enhances the visibility of the gap at the Dirac point and makes this relation between driving field strength and frequency desirable. 
We rewrite Eq.~\ref{edcond} to find the driving frequency that is necessary for a given field strength $\ed$ to have the gap $\Delta_0$ be equal to $m$ times the driving frequency $\od$.
It is 
\begin{equation}
    \Omega_m = \left(\frac{m^2}{4}+\frac{m}{2}\right)^{-\frac{1}{4}}\sqrt{e v_F\hbar^{-1}\ed}.
    \label{wdcond}
\end{equation}
The driving frequencies $\od=\Omega_m$ have the highest distinguishability and are indicated in Fig.~\ref{wdscan} (b) as vertical dashed lines. 
Additionally, at the frequency $\od=\Omega_1=(\frac{4}{3})^{\frac{1}{4}}\sqrt{e v_F \hbar^{-1} \ed}$, the energy of the first Floquet replica at $\Delta_0/2+\hbar\od$ is minimized.
This point denotes a regime that is well suited for trARPES probing and the conditions for resolvability simplify to $\gamma<(\frac{4}{3})^\frac{1}{4}\sqrt{e v_F \hbar^{-1} \ed}$.
For $\gamma=75\si{\tera\hertz}$ and $\ed\approx 20\si{\mega\volt\per\meter}$, this suggests a driving frequency close to $\od=\Omega_1\approx2\pi\times 29.8 \si{\tera\hertz}\approx 123\si{\milli\electronvolt}\hbar^{-1}$
or $\od=\Omega_2\approx 2\pi\times 23 \si{\tera\hertz}\approx 96\si{\milli\electronvolt}\hbar^{-1}$.

To demonstrate the steady state that emerges for driving at the frequencies $\od = \Omega_1$, $\Omega_2$, and $\Omega_3$, we show the electron distribution $n({\bf k}, \omega)$ in Fig.~\ref{fullnkw}.
This expands on the steady state behavior of the electron distribution $n({\bf k}=0, \omega)$ that we displayed in Fig.~\ref{wdscan}.
We choose the driving field strength $E_\mathrm{d}=20\si{\mega\volt\per\meter}$.
In the vicinity of the Dirac point, band occupations of the lower Floquet replica are suppressed.
The Floquet replica with sizeable electron occupation are two upper and two lower effective bands at the Dirac point, which are the four bands shown in Fig.~\ref{wdscan}.
As the gap $\Delta_0$ increases, the population is predominantly distributed among the two lower bands.
Away from the Dirac point, for non-zero ${\bf k}$, the additional Floquet bands have sizeable electron occupation and are visible in Fig.~\ref{fullnkw}.

Having pointed out the regime that we propose to use to detect the energy gap at the Dirac point in terms of the driving field strength and the driving frequency, we now present the time resolved response of the system.
Fig.~\ref{delayscans} shows the electron distribution at the Dirac point $n({\bf k}=0, \omega)$ as a function of the pulse delay time $\Delta t$ for the same driving field strength and driving frequencies as Fig.~\ref{fullnkw}, i.e. $\Omega_1$, $\Omega_2$ and $\Omega_3$ for $\ed=20\si{\mega\volt\per\meter}$.
This gives an estimate of the time-resolved Floquet-Bloch band occupations at the Dirac point.
The dashed lines indicate the corresponding Floquet energies expected from static driving field strengths given by the drive pulse at the delay time $\Delta t$, i.e. 
\begin{equation}
\epsilon(\Delta t) = \pm \sqrt{
    \exp\{ - \frac{\Delta t^2}{ \tau_\mathrm{d}^2} 8 \ln(2) \}
    (\frac{ev_F\ed}{\od})^2 +(\frac{\hbar\od}{2})^2}\pm\frac{\hbar\od}{2}.
\end{equation}
The electron distributions that we show in Fig.~\ref{delayscans} are close to the instantaneous steady state distribution, for this value of $\gamma$. Deviations from the instantaneous steady state distribution manifest themselves as features that are asymmetric during the pulse rise and pulse decay. For this choice of $\gamma$ and of the pulse lengths, these features are small. 

One common phenomenon that obscures the results of trARPES is laser-assisted photoemission (LAPE) \cite{Saathoff8}.
The essentially free photoelectrons emitted in a trARPES experiment respond to the drive pulse with driving frequency $\omega_\mathrm{d}$.
This may result in the photoelectron energy being shifted by one unit of the photon energy $\pm\hbar\omega_\mathrm{d}$.
These energy shifts are detected in trARPES measurements as band replica, whose similarities to Floquet replica might hinder identifying the signatures of Floquet physics unambiguously.
However, in contrast to Floquet replica, these LAPE replica are not related to band gaps \cite{ARPESreview}. 
The magnitude of the light-induced Floquet band gaps is tunable via the field strength $\ed$, see Eq.~\ref{delta0}.  
We propose to use this tunability to distinguish the LAPE and the Floquet replica. 
More specifically, the Floquet replica at the Dirac point are at $\pm \Delta_0/2$ and $\pm(\Delta_0/2+\hbar\od)$ as we show in Fig.~\ref{fullnkw}.
The monotonous behavior of the Dirac gap $\Delta_0$ makes it possible to distinguish between LAPE and Floquet replica.

The Floquet-Bloch bands resolved in $n({\bf k},\omega)$ are broadened due to dissipation.
In addition they are Fourier broadened with the probe pulse length. 
The combined result is a Voigt profile of approximate width 
\begin{equation}
\Gamma \approx \frac{\gamma}{2} + \sqrt{\frac{\gamma^2}{4} + \frac{4}{\tau_\mathrm{p}^2}}
\end{equation}
with $\gamma=\gamma_\pm/2+2\gamma_z+\gamma_\mathrm{bg}$. For our specified values this is $\Gamma=80\si{\tera\hertz}=(12.5\si{\femto\second})^{-1}\approx53\si{\milli\electronvolt}\hbar^{-1}$.
In order to successfully resolve the effective bands in $n({\bf k},\omega)$ it is crucial that the bands gaps are large compared to $\Gamma$. 
For probe pulses long enough such that their contribution to $\Gamma$ can be neglected, the broadening is $\gamma$ due to intrinsic dissipation.
Reducing $\gamma$ can be achieved by using cleaner graphene samples with higher mobility, which is technologically challenging. 

Furthermore, trARPES experiments are in general limited by a Gaussian pulse energy resolution of the order \cite{ARPESreview}
\begin{equation}
\Delta E \approx \tau_\mathrm{p}^{-1}1825\si{\milli\electronvolt}\si{\femto\second}.
\label{deltaE}
\end{equation}
The energy resolution of the measurement has to exceed the band gap $\Delta_0$, the Floquet replica spacing $\od$ and the Floquet-Bloch band Voigt width $\Gamma$.
These requirements are realistically achieved in the proposed regime of $\Delta_0>\hbar\od$. 
For instance, fulfilling the resolvability conditions $\Gamma \ll \omega_\mathrm{d}$ and $\Delta E \ll \hbar\omega_\mathrm{d}$ is not a necessity for identifying signatures of Floquet physics for this regime.

To determine the minimal probe length that is necessary to achieve an energy resolution equal to the gap, we insert the expression of the gap $\Delta_0$ at the Dirac point into Eq.~\ref{deltaE}.
It is 
\begin{equation}
\tau^\mathrm{min}_\mathrm{p}=\frac{1825\si{\milli\electronvolt}\si{\femto\second}}{\sqrt{4e^2v_F^2 E_\mathrm{d}^2/\omega_\mathrm{d}^2+\hbar^2\omega_\mathrm{d}^2}-\hbar\omega_\mathrm{d}}.
\end{equation}
An energy resolution several times better is necessary to clearly identify the Floquet-Bloch bands, which corresponds to probe pulse lengths several times larger than the minimal length, e.g. $\tau_\mathrm{p}\approx10\tau_\mathrm{p}^\mathrm{min}$.

In conclusion, we have pointed out a realistic regime for the detection of the light-induced topological gap in graphene via time- and angle-resolved photoelectron spectroscopy (trARPES).
Our proposed regime adresses the limitations of band broadening, energy- and momentum-resolution, and intrinsic limitations of the detection method for realistic estimates of dissipative processes.
We find that these limitations are overcome by increasing the driving field strength and decreasing the driving frequency so that the energy difference between finitely occupied Floquet-Bloch bands is larger than the Floquet zone.
The time-scales associated with the dissipative processes set the limits of this regime. 
On one hand, the driving frequency has to be large enough such that many driving cycles occur during one characteristic timescale of the dissipation.
On the other, decreasing the driving frequency increases the gap size at the Dirac point, which has to exceed the band broadening.
As the gap becomes larger than multiples of the driving frequency, limitations such as band broadening and inherent energy resolutions no longer obstruct the identification of signatures of Floquet physics.
This regime also allows undesired laser-assisted photoemission (LAPE) replica in trARPES measurements to be unambiguously identified as such at the Dirac point and to be clearly distinguished from the Floquet replica. 
The detection of Floquet bands via trARPES would consititute a profound insight in light-driven solids, that complements the measurements of transport reported in Ref.~\cite{mciver}, and thereby advances the field of optical control of solids.

\begin{acknowledgements}
We thank Kai Rossnagel, James McIver and Gregor Jotzu for very insightful discussions.
This work is funded by the Deutsche Forschungsgemeinschaft (DFG, German Research Foundation) -- SFB-925 -- project 170620586,
and the Cluster of Excellence "Advanced Imaging of Matter" (EXC 2056), Project No. 390715994.
\end{acknowledgements}

\bibliography{lit}
\bibliographystyle{unsrt}
\end{document}